\documentclass[10pt]{article}

\usepackage{graphics}

\newcommand{\gea}{\raisebox{-.3ex}{\small $ \
\stackrel{\textstyle >}{\sim} $ }}
\newcommand{\bbox}[1]{\mbox{\boldmath $#1$}}
\newcommand{\beq}{\begin{equation}}
\newcommand{\eeq}{\end{equation}}
\newcommand{\beqa}{\begin{eqnarray}}
\newcommand{\eeqa}{\end{eqnarray}}

\begin{document}
\title{\bf Chiral Symmetry and the \\ Nucleon-Nucleon Interaction\thanks{Chapter 14
of: {\it From Nuclei to Stars---Festschrift in Honor of Gerald E. Brown}, edited by
Sabine Lee (World Scientific, Singapore, 2011), p.~317-343.}}
\author{R. Machleidt\thanks{Electronic address: machleid@uidaho.edu}\\
Department of Physics, University of Idaho, \\ Moscow, ID 83843, U. S. A.
\and
D. R. Entem\thanks{Electronic address: entem@usal.es}\\
Grupo de Fisica Nuclear and IUFFyM, University of Salamanca, \\ E-37008 Salamanca, Spain}

\date{\today}

\maketitle


\begin{abstract}
We summarize the current status of our understanding of nuclear forces based upon 
chiral symmetry---an idea that was advocated by Gerry Brown already several decades ago.
\end{abstract}

\maketitle

\section{Introduction}

Forty years ago, Gerry Brown published the Comment
``Isn't it Time to Calculate the Nucleon-Nucleon Force''~\cite{Bro70}, and
more than 30 years ago, he wrote
a book chapter entitled ``Chiral Symmetry and the Nucleon-Nucleon
Interaction''\cite{Bro79}.
In fact, as early as 1968, he had published, together with two co-workers,
a paper~\cite{BGG68} on three-nucleon forces, where the consequences of chiral symmetry were fully
exploited. Nevertheless, it should take a few more {\it decades} until the problem of the nucleon-nucleon ($NN$)
interaction was really solved taking chiral symmetry consistently into
account---and this fact proves the very advanced nature
of Gerry's ideas concerning nuclear forces. It is the purpose of this contribution to summarize
where we are today in our understanding of nuclear forces, based upon concepts that Gerry
advocated already many decades ago.

But let's first recall more of the interesting history 
of the idea of chiral symmetry.
The modern understanding is that this symmetry arises because the up and
down quarks happen to have relatively small masses.
However, chiral symmetry and its significance for low-energy hadron (pion)
physics was discovered long before QCD.
In 1960, based upon concepts proposed by Schwinger~\cite{Sch57},
Gell-Mann and Levy~\cite{GL60} developed the sigma model, which is
a linear realization of chiral symmetry.\footnote{For a pedagogical introduction into chiral symmetry
and the sigma model, see~\cite{EW88}.}
One major problem researchers had been struggling with in the 1950's was that the pion-nucleon
scattering length came out two orders of magnitude too
large when the (renormalizable) pseudo-scalar ($\gamma_5$) $\pi N$ interaction was used.
This unrealistic prediction was due to very large contributions from
virtual anti-nucleon states (the so-called ``pair terms'' or ``Z-graphs''). Similar problems
occurred in the $2\pi$-exchange contribution to the $NN$ interaction. 
In the sigma model, the large pair terms are cancelled by processes
involving the (fictitious) $\sigma$ boson. In this way, the linear sigma model
demonstrates how imposing chiral invariance fixes the problem with low-energy
$\pi$-$N$ scattering. However, the fictitious character of the $\sigma$ particle as well as 
the reliance on the perfect
cancelation of huge terms are uncomfortable features.
In 1967, motivated by the current algebra approach to soft pion physics,
Weinberg~\cite{Wei67} worked out what has become known as the non-linear sigma model, which
does not include a $\sigma$ anymore and has pions and nucleons interact via pseudo-vector
(derivative, $\gamma_5 \gamma^\mu \partial_\mu$) 
coupling besides a new (non-linear) $\pi\pi NN$ term also involving a derivative
(``Weinberg-Tomozawa term''~\cite{Wei66,Tom66}).
The derivative (equivalent to momentum) guarantees that the interaction vanishes when the momentum goes to zero providing a natural explanation for the weakness of the interaction by
soft pions  which does not rely on the cancellation of large terms. 
Following suggestions by Schwinger, Weinberg~\cite{Wei68} developed, soon after, 
a general theory of non-linear realizations
of chiral symmetry, which was further generalized in an elegant way by
Callan, Coleman, Wess, and Zuimino~\cite{CCWZ}.

However, ideas were needed for how
to implement chiral symmetry consistently in the theory of pionic and nuclear interactions
and how to deal with renormalization, since the derivative coupling is not renormalizable
in the conventional sense.
In his contribution to the `Festschrift' in honor of Schwinger of 1979~\cite{Wei79,Wei09}, Weinberg proposed
to consider the most general possible Langrangian including all higher-derivative 
terms that are consistent
with chiral symmetry (besides the other commonly assumed symmetry principles).
For this theory to be manageable, one needs to assume some sort of perturbative
expansion such that only a finite number of terms contribute at a given order.
This expansion is provided by powers of small external momenta over the chiral symmetry
breaking scale, $\Lambda_\chi \sim 1$ GeV. The higher-derivative terms supply the counterterms
that make possible an order-by-order renormalization, which is the appropriate renormalization 
procedure for an effective field theory.
Weinberg's suggestions were soon picked up  
by Gasser, Leutwyler, and associates who worked out, to one loop, the cases of 
$\pi\pi$~\cite{GL84} and $\pi N$ scattering~\cite{GSS88} with great success.

But there was still the problem of the nuclear force which is more difficult, 
since nuclear interactions do not vanish in the chiral limit ($q\rightarrow 0$; $m_{u/d}, m_\pi \rightarrow 0$)
and require a non-perturbative
treatment because of the existence of nuclear bound states.
In a series of papers published around 1990~\cite{Wei90,Wei91,Wei92}, 
Weinberg picked up the nuclear force issue and suggested to calculate the $NN$ potential
perturbatively in the chiral expansion and then iterate it to all orders in a 
Schroedinger or Lippmann-Schwinger
equation to obtain the nuclear amplitude. 
Here, the introduction of four-nucleon contact terms is crucial for renormalization.

Following the Weinberg proposal, pioneering
work was performed by Ord\'o\~nez, Ray, and
van Kolck \cite{ORK94,ORK96} who applied time-ordered perturbation theory to
construct a $NN$ potential up to next-to-next-to-leading order (NNLO).
The results were encouraging and nuclear EFT quickly developed into one of the most popular
branches of modern nuclear physics.
The Munich group used covariant perturbation theory and dimensional regularization
to calculated the perturbative
$NN$ amplitude without~\cite{KBW97} and 
with $\Delta(1232)$-isobar degrees of freedom~\cite{KGW98} at NNLO.
Besides this, the Munich group worked out important loop contributions of
higher order~\cite{Kai00a,Kai00b,Kai01,Kai01a,Kai01b}.
A relativistic approach was also taken by the Brazil group~\cite{RR94,RR03}.
The Bochum-J\"ulich group devised
a method of unitarity transformations to eliminate the energy-dependence of 
time-ordered perturbation theory amplitudes and calculated 
the $NN$ potentials up to NNLO~\cite{EGM98,EGM00}.
The Idaho group managed to construct a chiral $NN$ potential
at next-to-next-to-next-to-leading order (N$^3$LO) and showed that only at
this order can one achieve the precision necessary for reliable few-nucleon
and nuclear structure calculations~\cite{EM02a,EM02,EM03,ME05}.
Progress extended beyond the $NN$ interaction, as nuclear many-body forces based 
upon chiral perturbation theory were also developed~\cite{Wei92,Kol94,Epe02b,IR07,Ber08}.

During the past decade or so,
chiral two-nucleon forces have been used in many microscopic calculations of nuclear reactions and structure~\cite{DF07,Cor02,Cor05,Cor10,NC04,FNO05,Var05,Kow04,DH04,Wlo05,Dea05,Gou06,Hag08,Hag10,FOS04,FOS09} and the combination of chiral two- and three-nucleon forces has been applied in
few-nucleon reactions~\cite{Epe02b,Erm05,Kis05,Wit06,Ley06,Ste07,KE07,Mar09,Kie10,Viv10},
structure of light- and medium-mass nuclei~\cite{Nog06,Nav07,Hag07,Ots09},
and nuclear and neutron matter~\cite{Bog05,HS09}---with a great deal of success.
The majority of nuclear structure calculations is nowadays based upon chiral forces.

This article is organized as follows.
In Section~\ref{sec_EFT}, we sketch the essential ideas of an EFT for low-energy
QCD. Section~\ref{sec_overview} provides an overview
on nuclear forces derived from chiral EFT. The two-nucleon force is then discussed in detail in Section~\ref{sec_NN}.
Many-body forces are the subject of Section~\ref{sec_manyNF}. 
Finally, Section~\ref{sec_concl} contains our conclusions.

\section{Low-energy QCD and effective field theory
\label{sec_EFT}}

Quantum chromodynamics (QCD) is the theory of strong interactions.
It deals with quarks, gluons and their interactions and is
part of the Standard Model of Particle Physics.
QCD is a non-Abelian gauge field theory
with color $SU(3)$ the underlying gauge group.
The non-Abelian nature of the theory has dramatic
consequences. While 
the interaction between colored objects is weak 
at short distances or high momentum transfer
(``asymptotic freedom'');
it is strong at long distances ($\gea 1$ fm) or low energies,
leading to the confinement of quarks into colorless
objects, the hadrons. Consequently, QCD allows for a 
perturbative analysis at large energies, whereas it is
highly non-perturbative in the low-energy regime.
Nuclear physics resides at low energies and
the force between nucleons is
a residual color interaction
similar to the van der Waals force between neutral molecules.
Therefore, in terms of quarks and gluons, the nuclear force
is a very complicated problem that, nevertheless, can be attacked
with brute computing power on a discretized, Euclidean space-time lattice
(known as lattice QCD).
Advanced lattice QCD calculations~\cite{Bea06,IAH07} are under way and will yield improving
results in the near future. However, since these calculations are very time-consuming
and expensive, they can only be used to check a few representative key-issues. For everyday
nuclear structure physics, a more efficient approach is needed. 

The efficient approach is an effective field theory.
For the development of an EFT, it is crucial to identify a separation of
scales. In the hadron spectrum, a large gap between the masses of
the pions and the masses of the vector mesons, like $\rho(770)$ and $\omega(782)$,
can clearly be identified. Thus, it is natural to assume that the pion mass sets the soft scale, 
$Q \sim m_\pi$,
and the rho mass the hard scale, $\Lambda_\chi \sim m_\rho$, also known
as the chiral-symmetry breaking scale.
This is suggestive of considering an expansion in terms of the soft scale over the hard scale,
$Q/\Lambda_\chi$.
Concerning the relevant degrees of freedom, we noticed already that,
for the ground state and the
low-energy excitation spectrum of
an atomic nucleus as well as for conventional nuclear
reactions,
quarks and gluons are ineffective degrees of freedom,
while nucleons and pions are the appropriate ones.
To make sure that this EFT is not just another phenomenology,
it must have a firm link with QCD.
The link is established by having the EFT observe
all relevant symmetries of the underlying theory.
This requirement is based upon a `folk theorem' by
Weinberg~\cite{Wei79}:
\begin{quote}
If one writes down the most general possible Lagrangian, including {\it all}
terms consistent with assumed symmetry principles,
and then calculates matrix elements with this Lagrangian to any given order of
perturbation theory, the result will simply be the most general possible 
S-matrix consistent with analyticity, perturbative unitarity,
cluster decomposition, and the assumed symmetry principles.
\end{quote}

Since the up and down quark masses are very small, one may assume
(approximate) chiral symmetry which, however, is broken in two ways.
It is explicitly broken, because the the up and down quark masses are not exactly zero.
Moreover, we are also faced with so-called spontaneous symmetry breaking.

A (continuous) symmetry is said to be {\it spontaneously
broken} if a symmetry of the Lagrangian 
is not realized in the ground state of the system.
There is evidence that the (approximate) chiral
symmetry of the QCD Lagrangian is spontaneously 
broken---for dynamical reasons of nonperturbative origin
which are not fully understood at this time.
The most plausible evidence comes from the hadron spectrum.
From chiral symmetry, one would naively expect the existence of 
degenerate hadron
multiplets of opposite parity, i.e., for any hadron of positive
parity one would expect a degenerate hadron state of negative 
parity and vice versa. However, these ``parity doublets'' are
not observed in nature. For example, take the $\rho$-meson which is
a vector meson of negative parity ($J^P=1^-$) and mass 
776 MeV. There does exist a $1^+$ meson, the $a_1$, but it
has a mass of 1230 MeV and, thus, cannot be perceived
as degenerate with the $\rho$. On the other hand, the $\rho$
meson comes in three charge states (equivalent to
three isospin states), the $\rho^\pm$ and the $\rho^0$,
with masses that differ by at most a few MeV. In summary,
in the hadron spectrum,
$SU(2)_V$ (isospin symmetry) is well observed,
while axial symmetry is broken:
$SU(2)_L\times SU(2)_R$ is broken down to $SU(2)_V$.

A spontaneously broken global symmetry implies the existence
of (massless) Goldstone bosons~\cite{Gol61,GSW62}. 
The Goldstone bosons are identified with the isospin
triplet of the (pseudoscalar) pions, 
which explains why pions are so light.
The pion masses are not exactly zero because the up
and down quark masses
are not exactly zero either.
Thus, pions are a truly remarkable species:
they reflect spontaneous as well as explicit symmetry
breaking.
Goldstone bosons interact weakly at low energy.
They are degenerate with the vacuum 
and, therefore, interactions between them must
vanish at zero momentum and in the chiral limit
($m_\pi \rightarrow 0$).

The next step is to build the most general
Lagrangian consistent with the (broken) symmetries discussed
above.
An elegant formalism for the construction of such Lagrangians
was developed by 
Callan, Coleman, Wess, and Zumino (CCWZ)~\cite{CCWZ}
who worked out
the group-theoretical foundations 
of non-linear realizations of chiral symmetry.
It is characteristic for these non-linear realizations that, whenever functions of the Goldstone bosons
appear in the Langrangian, they are always accompanied with at least one
space-time derivative.

As discussed, the relevant degrees of freedom are
pions (Goldstone bosons) and nucleons.
Since the interactions of Goldstone bosons must
vanish at zero momentum transfer and in the chiral
limit ($m_\pi \rightarrow 0$), the low-energy expansion
of the Lagrangian is arranged in powers of derivatives
and pion masses.
The hard scale is the chiral-symmetry breaking
scale, $\Lambda_\chi \approx 1$ GeV. Thus, the expansion is in terms
of powers of $Q/\Lambda_\chi$ where $Q$ is a (small) momentum
or pion mass.
This is chiral perturbation theory (ChPT).

The effective Lagrangian can formally be written as,
\begin{equation}
{\cal L_{\rm eff}} 
=
{\cal L}_{\pi\pi} 
+
{\cal L}_{\pi N} 
+
{\cal L}_{NN} 
 + \, \ldots \,,
\end{equation}
where ${\cal L}_{\pi\pi}$
deals with the dynamics among pions, 
${\cal L}_{\pi N}$ 
describes the interaction
between pions and a nucleon,
and
${\cal L}_{NN}$
provides contact interactions between two nucleons.
The individual Lagrangians are organized as follows:
\begin{equation}
{\cal L}_{\pi\pi} 
 = 
{\cal L}_{\pi\pi}^{(2)} 
 + {\cal L}_{\pi\pi}^{(4)} 
 + \ldots ,
\end{equation}
\begin{equation}
{\cal L}_{\pi N} 
= 
{\cal L}_{\pi N}^{(1)} 
+
{\cal L}_{\pi N}^{(2)} 
+
{\cal L}_{\pi N}^{(3)} 
+ \ldots ,
\end{equation}
and
\begin{equation}
{\cal L}_{NN} 
= 
{\cal L}_{NN}^{(0)} 
+
{\cal L}_{NN}^{(2)} 
+
{\cal L}_{NN}^{(4)} 
+ \ldots ,
\end{equation}
where the superscript refers to the number of derivatives or 
pion mass insertions (chiral dimension)
and the ellipsis stands for terms of higher dimensions.
See Ref.~\cite{EM02} for a concise summary of the explicit expressions
for these Lagrangians.

\section{Nuclear forces from chiral EFT: Overview
\label{sec_overview}}

\subsection{Chiral perturbation theory and power counting}
As discussed,
effective field theories (EFTs) are defined in terms of effective Langrangians which
are given by an infinite series of terms with increasing number of derivatives
and/or nucleon fields, with the dependence of each term on the pion field 
prescribed by the rules of broken chiral symmetry.
Applying this Lagrangian to a particular process, an unlimited number of Feynman 
graphs can be generated. Therefore,
we need a scheme that makes the theory manageable and calculable.
This scheme
which tells us how to distinguish between large
(important) and small (unimportant) contributions
is chiral perturbation theory (ChPT), and
determining the power $\nu$ of the expansion
has become known as power counting.

Nuclear potentials are defined as sets of irreducible
graphs up to a given order.
The power $\nu$ of a few-nucleon diagram involving $A$ nucleons
is given in terms of naive dimensional analysis by:
\begin{equation} 
\nu = -2 +2A - 2C + 2L 
+ \sum_i \Delta_i \, ,  
\label{eq_nu} 
\end{equation}
with
\begin{equation}
\Delta_i  \equiv   d_i + \frac{n_i}{2} - 2  \, ,
\label{eq_Deltai}
\end{equation}
where $C$ denotes the number of separately connected pieces and
$L$ the number of loops in the diagram;
$d_i$ is the number of derivatives or pion-mass insertions and $n_i$ the number of nucleon fields 
(nucleon legs) involved in vertex $i$; the sum runs over all vertices contained
in the diagram under consideration.
Note that $\Delta_i \geq 0$
for all interactions allowed by chiral symmetry.
For an irreducible $NN$ diagram (``two-nucleon force'', $A=2$, $C=1$),
Eq.~(\ref{eq_nu}) collapses to
\begin{equation} 
\nu =  2L + \sum_i \Delta_i \, .  
\label{eq_nunn} 
\end{equation}

In summary, the chief point of the ChPT expansion is that,
at a given order $\nu$, there exists only a finite number
of graphs. This is what makes the theory calculable.
The expression $(Q/\Lambda_\chi)^{\nu+1}$ provides a rough estimate
of the relative size of the contributions left out and, thus,
of the accuracy at order $\nu$.
In this sense, the theory can be calculated to any
desired accuracy and has
predictive power.

\subsection{The hierarchy of nuclear forces}
\begin{figure}[t]\centering
\vspace*{-0.5cm}
\scalebox{0.65}{\includegraphics{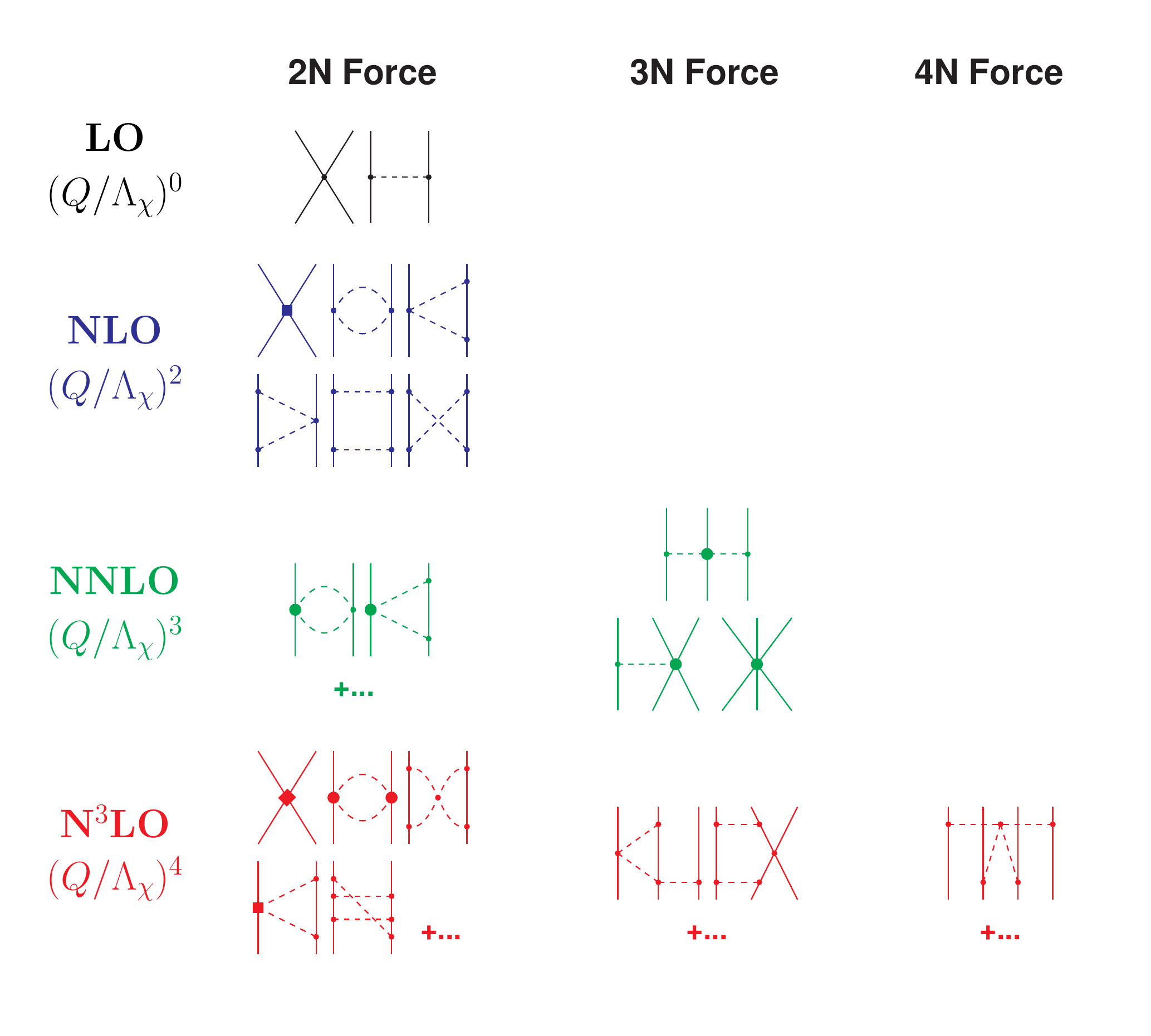}}
\vspace*{-0.75cm}
\caption{Hierarchy of nuclear forces in ChPT. Solid lines
represent nucleons and dashed lines pions. 
Small dots, large solid dots, solid squares, and solid diamonds
denote vertices of index $\Delta= \, $ 0, 1, 2, and 4, respectively. 
Further explanations are
given in the text.}
\label{fig_hi}
\end{figure}

Chiral perturbation theory and power counting
imply that nuclear forces emerge as a hierarchy
controlled by the power $\nu$, Fig.~\ref{fig_hi}.

In lowest order, better known as leading order (LO, $\nu = 0$), 
the $NN$ amplitude
is made up by two momentum-independent contact terms
($\sim Q^0$), 
represented by the 
four-nucleon-leg graph
with a small-dot vertex shown in the first row of 
Fig.~\ref{fig_hi},
and
static one-pion exchange (1PE), second
diagram in the first row of the figure.
This is, of course, a rather crude approximation
to the two-nucleon force (2NF), but accounts already for some
important features.
The 1PE provides the tensor force,
necessary to describe the deuteron, and it explains
$NN$ scattering in peripheral partial waves of very high
orbital angular momentum. At this order, the two contacts 
which contribute only in $S$-waves provide
the short- and intermediate-range interaction which is somewhat
crude.

In the next order,
$\nu=1$, all contributions vanish due to parity
and time-reversal invariance.

Therefore, the next-to-leading order (NLO) is $\nu=2$.
Two-pion exchange (2PE) occurs for the first time
(``leading 2PE'') and, thus, the creation of
a more sophisticated
description of the intermediate-range interaction
is starting here. 
Since the loop involved in each pion-diagram implies
already $\nu=2$ [cf.\ Eq.~(\ref{eq_nunn})],
the vertices must have $\Delta_i = 0$.
Therefore, at this order, only the lowest order
$\pi NN$ and $\pi \pi NN$ vertices are allowed which
is why the leading 2PE is rather weak.
Furthermore, there are 
seven contact terms of 
${\cal O}(Q^2)$, 
shown by
the four-nucleon-leg graph with a solid square,
 which contribute
in $S$ and $P$ waves. The operator structure of these
contacts include a spin-orbit term besides central,
spin-spin, and tensor terms. Thus, essentially all spin-isospin
structures necessary to describe the two-nucleon
force phenomenologically have been generated at this order.
The main deficiency at this stage of development 
is an insufficient intermediate-range attraction.

This problem is finally fixed at order three 
($\nu=3$), next-to-next-to-leading order (NNLO).
The 2PE involves now the two-derivative
$\pi\pi NN$ seagull vertices (proportional to
the $c_i$ LECs) denoted by a large solid dot
in Fig.~\ref{fig_hi}.
These vertices represent correlated 2PE
as well as intermediate $\Delta(1232)$-isobar contributions.
It is well-known from the meson phenomenology of 
nuclear forces~\cite{MHE87,Lac80}
that these two contributions are crucial
for a realistic and quantitative 2PE model.
Consequently, the 2PE now assumes a realistic size
and describes the intermediate-range attraction of the
nuclear force about right. Moreover, first relativistic 
corrections come into play at this order.
There are no new contacts.

The reason why we talk of a hierarchy of nuclear forces is that 
two- and many-nucleon forces are created on an equal footing
and emerge in increasing number as we go to higher and higher orders.
At NNLO, the first set of
nonvanishing three-nucleon forces (3NF) occur~\cite{Kol94,Epe02b},
cf.\ column `3N Force' of
Fig.~\ref{fig_hi}. 
In fact, at the previous order, NLO,
irreducible 3N graphs appear already, however,
it has been shown by Weinberg~\cite{Wei92} and 
others~\cite{Kol94,YG86,CF86} that these diagrams all cancel.
Since nonvanishing 3NF contributions happen first
at order 
$(Q/\Lambda_\chi)^3$, 
they are very weak as compared to 2NF which start at
$(Q/\Lambda_\chi)^0$.

More 2PE is produced at $\nu =4$, next-to-next-to-next-to-leading
order (N$^3$LO), of which we show only a few symbolic diagrams in 
Fig.~\ref{fig_hi}. 
Two-loop 2PE
graphs show up for the first time and so does
three-pion exchange (3PE) which necessarily involves
two loops.
3PE was found to be negligible at this order~\cite{Kai00a,Kai00b}.
Most importantly, 15 new contact terms $\sim Q^4$
arise and are represented 
by the four-nucleon-leg graph with a solid diamond.
They include a quadratic spin-orbit term and
contribute up to $D$-waves.
Mainly due to the increased number of contact terms,
a quantitative description of the
two-nucleon interaction up to about 300 MeV
lab.\ energy is possible, 
at N$^3$LO 
(for details, see below).
Besides further 3NF,
four-nucleon forces (4NF) start
at this order. Since the leading order 4NF 
come into existence one
order higher than the leading 3NF, 4NF are weaker
than 3NF.
Thus, ChPT provides a straightforward explanation for
the empirically known fact that 2NF $\gg$ 3NF $\gg$ 4NF
\ldots.

\section{The nucleon-nucleon interaction
\label{sec_NN}}

\subsection{Definition of the chiral $NN$ potential}

The previous section has provided us with an overview. In this section, we will now 
discuss the $NN$ interaction in more detail.
In terms of naive dimensional analysis or ``Weinberg counting'',
the various orders of the irreducible graphs which define the chiral $NN$ potential 
are given by:
\beqa
V_{\rm LO} & = & 
V_{\rm ct}^{(0)} + 
V_{1\pi}^{(0)} 
\label{eq_VLO}
\\
V_{\rm NLO} & = & V_{\rm LO} +
V_{\rm ct}^{(2)} + 
V_{1\pi}^{(2)} +
V_{2\pi}^{(2)} 
\label{eq_VNLO}
\\
V_{\rm NNLO} & = & V_{\rm NLO} +
V_{1\pi}^{(3)} + 
V_{2\pi}^{(3)} 
\label{eq_VNNLO}
\\
V_{{\rm N}^3{\rm LO}} & = & V_{\rm NNLO} +
V_{\rm ct}^{(4)} +
V_{1\pi}^{(4)} +  
V_{2\pi}^{(4)} +
V_{3\pi}^{(4)} 
\label{eq_VN3LO}
\eeqa
where 
the superscript denotes the order $\nu$ of the low-momentum
expansion.
LO stands for leading order, NLO for next-to-leading
order, etc..
Contact potentials carry the subscript ``ct'' and
pion-exchange potentials can be identified by an
obvious subscript.

The one-pion exchange (1PE) potential reads
\begin{equation}
V_{1\pi} ({\vec p}~', \vec p) = - 
\frac{g_A^2}{4f_\pi^2}
\: 
\bbox{\tau}_1 \cdot \bbox{\tau}_2 
\:
\frac{
\vec \sigma_1 \cdot \vec q \,\, \vec \sigma_2 \cdot \vec q}
{q^2 + m_\pi^2} 
\,,
\label{eq_1peci}
\end{equation}
where ${\vec p}~'$ and $\vec p$ designate the 
final and initial nucleon momenta 
in the center-of-mass system and
$\vec q \equiv {\vec p}~' - \vec p$  is the 
momentum transfer;
$\vec \sigma_{1,2}$ and $\bbox{\tau}_{1,2}$ are 
the spin and isospin 
operators of nucleon 1 and 2;
$g_A$, $f_\pi$, and $m_\pi$
denote axial-vector coupling constant, the pion decay constant,
and the pion mass, respectively.
Since higher order corrections contribute only to mass and
coupling constant renormalizations and since, on shell,
there are no relativistic corrections, the on-shell
1PE has the form Eq.~(\ref{eq_1peci}) up to all orders.

Multi-pion exchange, which starts at NLO and continues through
all higher orders, involves
divergent loop integrals that need to be regularized.
An elegant way to do this is dimensional regularization
which 
(besides the main nonpolynomial result) 
typically generates polynomial terms with coefficients
that are, in part, infinite or scale dependent~\cite{KBW97}.
One purpose of the contacts is
to absorb all infinities and scale dependencies and make
sure that the final result is finite and scale independent.
This is the renormalization of the perturbatively calculated
$NN$ amplitude (which, by definition, is the ``$NN$ potential'').
It is very similar to what is done in the ChPT calculations
of $\pi\pi$ and $\pi N$ scattering, namely, a renormalization
order by order, which is the method of choice for any EFT.
Thus, up to this point, the calculation fully meets the
standards of an EFT and there are no problems.
The perturbative $NN$ amplitude can be used to make model
independent predictions for peripheral partial waves~\cite{KBW97,KGW98,EM02}.
A concise summary of the explicit expressions for the $2\pi$ exchange contributions
up to order N$^3$LO can be found in Ref.~\cite{EM02}.

\subsection{Regularization and renormalization
\label{sec_reno}}
For calculations of the structure of nuclear few and many-body systems,
the lower partial waves are the most important ones. The fact that
in $S$ waves we have large scattering lengths and shallow (quasi)
bound states indicates that these waves need to be treated nonperturbatively.
Following Weinberg's prescription~\cite{Wei90}, this is accomplished by
inserting the potential $V$ into the Lippmann-Schwinger (LS) equation:
\begin{equation}
 {T}({\vec p}~',{\vec p})= {V}({\vec p}~',{\vec p})+
\int d^3p''\:
{V}({\vec p}~',{\vec p}~'')\:
\frac{M_N}
{{ p}^{2}-{p''}^{2}+i\epsilon}\:
{T}({\vec p}~'',{\vec p}) \,,
\label{eq_LS}
\end{equation}
where $M_N$ denotes the nucleon mass.

In general, the integral in
the LS equation is divergent and needs to be regularized.
One way to do this is  by
multiplying $V$
with a regulator function
\begin{equation}
{ V}(\vec{ p}~',{\vec p}) 
\longmapsto
{ V}(\vec{ p}~',{\vec p})
\;\mbox{\boldmath $e$}^{-(p'/\Lambda)^{2n}}
\;\mbox{\boldmath $e$}^{-(p/\Lambda)^{2n}}
\label{eq_regulator} \,.
\end{equation}
Typical choices for the cutoff parameter $\Lambda$ that
appears in the regulator are 
$\Lambda \approx 0.5 \mbox{ GeV} \ll \Lambda_\chi \approx 1$ GeV.

It is pretty obvious that results for the $T$-matrix may
depend sensitively on the regulator and its cutoff parameter.
This is acceptable if one wishes to build models.
For example, the meson models of the past~\cite{Mac89,MHE87}
always depended sensitively on the choices for the
cutoff parameters which, in fact,
were important for the fit of the $NN$ data.
However, the EFT approach wishes to be fundamental
in nature and not just another model.

In field theories, divergent integrals are not uncommon and methods have
been developed for how to deal with them.
One regulates the integrals and then removes the dependence
on the regularization parameters (scales, cutoffs)
by renormalization. In the end, the theory and its
predictions do not depend on cutoffs
or renormalization scales.

So-called renormalizable quantum field theories, like QED,
have essentially one set of prescriptions 
that takes care of renormalization through all orders. 
In contrast, 
EFTs are renormalized order by order. 

As discussed, the renormalization of {\it perturbative}
EFT calculations is not a problem. {\it The problem
is nonperturbative renormalization.}
This problem typically occurs in {\it nuclear} EFT because
nuclear physics is characterized by bound states which
are nonperturbative in nature.
EFT power counting may be different for nonperturbative processes as
compared to perturbative ones. Such difference may be caused by the infrared
enhancement of the reducible diagrams generated in the LS equation.

Weinberg's implicit assumption~\cite{Wei90,Wei09} was that the counterterms
introduced to renormalize the perturbatively calculated
potential, based upon naive dimensional analysis (``Weinberg counting''),
are also sufficient to renormalize the nonperturbative
resummation of the potential in the LS equation.
In 1996, Kaplan, Savage, and Wise (KSW)~\cite{KSW96}
pointed out that there are problems with the Weinberg scheme
if the LS equation is renormalized 
by minimally-subtracted dimensional regularization.
This criticism resulted in a flurry of publications on
the renormalization of the nonperturbative
$NN$ problem. See Ref.~\cite{ME10}
for a thorough discussion, the bottom line of which is:

Crucial for an EFT are regulator independence (within the range of validity
of the EFT) and a power counting scheme that allows for order-by-order
improvement with decreasing truncation error.
The purpose of renormalization is to achieve this regulator independence while maintaining
a functional power counting scheme.
After the comprehensive tries and errors of the past, it appears that there are two renormalization
schemes which have the potential to achieve the above goals and, therefore, should be investigated 
systematically in the near future.

In {\it scheme one}, the LO calculation is conducted nonperturbatively
(with $\Lambda \rightarrow \infty$ as in~\cite{NTK05})
and subleading orders are added perturbatively in distorted wave Born approximation.
Valderrama has started this in $S$ waves~\cite{Val09}, but results in higher
partial waves are needed to fully assess this approach.
Even though at this early stage any judgement is speculative, we take the liberty to predict
that this approach will be only of limited success and utility---for the following reasons.
First, it will probably require about twice as many counterterms as Weinberg counting
and, therefore, will have less predictive power. Second, this scheme may converge badly,
because the largest portion of the nuclear force, namely, the intermediate-range
attraction appears at NNLO. Third, as discussed in~\cite{Mac09}, this force may be problematic
(and, therefore, impractical) in applications in nuclear few- and many-body systems, 
because of a pathologically strong tensor force that will cause bad convergence of energy and wave functions. Finally, in the work that has been conducted so far within this
scheme by Valderrama,
it is found that only rather soft cutoffs can be used.

The latter point (namely, soft cutoffs) suggests that one may then as well
conduct the calculation nonperturbatively at all orders (up to N$^3$LO) using Weinberg counting, 
which is no problem with soft cutoffs. This is {\it scheme two} that we propose to
investigate systematically. 
In the spirit of Lepage~\cite{Lep97}, the cutoff independence should be examined
for cutoffs below the hard scale and not beyond. Ranges of cutoff independence within the
theoretical error are to be identified using `Lepage plots'.
A very systematic investigation of this kind does not exist at this time and is therefore needed.
However, there is comprehensive circumstantial
evidence from the numerous chiral $NN$ potentials constructed over the past 
decade~\cite{ORK96,EGM00,EM02,EM03,ME05,EGM05}
indicating that this investigation will most likely be a success
(cf.\ also Fig.~\ref{fig_phn3lo1}, below).
The potentials discussed in the following section are all based upon Weinberg counting.
 
 \begin{table}[t]
\caption{Columns three and four show the
$\chi^2$/datum for the reproduction of the 1999 $np$ 
database~\cite{note2} (subdivided into energy intervals)
by families of $np$ potentials at NLO and NNLO constructed by the
Juelich group~\cite{EGM04}.
The $\chi^2$/datum is stated in terms of ranges
which result from a variation of the cutoff parameters
used in the regulator functions.
The values of these cutoff parameters in units of MeV are given in 
parentheses.
$T_{\rm lab}$ denotes the kinetic energy of the incident neutron
in the laboratory system.
\label{tab_chi2a}}
\smallskip
\begin{tabular*}{\textwidth}{@{\extracolsep{\fill}}cccc}
\hline 
\hline 
\noalign{\smallskip}
 $T_{\rm lab}$ bin &  \# of $np$ & 
\multicolumn{2}{c}{\it --- Juelich $np$ potentials --- }\\
 (MeV) 
 & data 
 & NLO
 & NNLO 
\\
 & & (550/700--400/500) & (600/700--450/500)
 \\
\hline 
\hline 
\noalign{\smallskip}
0--100&1058&4--5&1.4--1.9\\ 
100--190&501&77--121&12--32\\ 
190--290&843&140--220&25--69\\ 
\hline 
\noalign{\smallskip}
0--290&2402&67--105&12--27
\\ 
\hline 
\hline 
\end{tabular*}
\end{table}

\begin{table}[t]
\small
\caption{\small Number of parameters needed for fitting the $np$ data
in phase-shift analysis and by a high-precision $NN$ potential
{\it versus} the total number of $NN$ contact terms of EFT based potentials 
to different orders. 
\label{tab_par}}
\small
\smallskip
\begin{tabular*}{\textwidth}{@{\extracolsep{\fill}}ccc|ccc}
\hline 
\hline 
\noalign{\smallskip}
     & Nijmegen     & CD-Bonn        & 
               \multicolumn{3}{c}{\it --- Contact Potentials --- }\\
     & partial-wave & high-precision & $Q^0$ & $Q^2$  & $Q^4$ \\
     & analysis~\cite{Sto93} & potential~\cite{Mac01} & 
                                 LO & NLO/NNLO  & N$^3$LO \\
\hline 
\hline 
\noalign{\smallskip}
$^1S_0$         & 3 & 4 & 1&2 & 4 \\
$^3S_1$         & 3 & 4 & 1&2 & 4 \\
\hline
\noalign{\smallskip}
$^3S_1$-$^3D_1$ & 2 & 2 & 0&1 & 3 \\
\hline
\noalign{\smallskip}
$^1P_1$         & 3 & 3 & 0&1 & 2 \\
$^3P_0$         & 3 & 2 & 0&1 & 2 \\
$^3P_1$         & 2 & 2 & 0&1 & 2 \\
$^3P_2$         & 3 & 3 & 0&1 & 2 \\
\hline
\noalign{\smallskip}
$^3P_2$-$^3F_2$ & 2 & 1 & 0&0 & 1 \\
\hline
\noalign{\smallskip}
$^1D_2$         & 2 & 3 & 0&0 & 1 \\
$^3D_1$         & 2 & 1 & 0&0 & 1 \\
$^3D_2$         & 2 & 2 & 0&0 & 1 \\
$^3D_3$         & 1 & 2 & 0&0 & 1 \\
\hline
\noalign{\smallskip}
$^3D_3$-$^3G_3$ & 1 & 0 & 0&0 & 0 \\
\hline
\noalign{\smallskip}
$^1F_3$         & 1 & 1 & 0&0 & 0 \\
$^3F_2$         & 1 & 2 & 0&0 & 0 \\
$^3F_3$         & 1 & 2 & 0&0 & 0 \\
$^3F_4$         & 2 & 1 & 0&0 & 0 \\
\hline
\noalign{\smallskip}
$^3F_4$-$^3H_4$ & 0 & 0 & 0&0 & 0 \\
\hline
\noalign{\smallskip}
$^1G_4$         & 1 & 0 & 0&0 & 0 \\
$^3G_3$         & 0 & 1 & 0&0 & 0 \\
$^3G_4$         & 0 & 1 & 0&0 & 0 \\
$^3G_5$         & 0 & 1 & 0&0 & 0 \\
\hline
\hline
\noalign{\smallskip}
Total         & 35  & 38 & 2&9 & 24 \\
\hline
\hline
\noalign{\smallskip}
\end{tabular*}
\end{table}

\subsection{Constructing quantitative chiral $NN$ potentials}

\subsubsection{What order?}

As discussed, the $NN$ potential can be calculated up to
various orders, cf.\ Eqs.~(\ref{eq_VLO})-(\ref{eq_VN3LO}),
and the accuracy increases as the order increases.
That triggers the obvious question:
To what order do we have to go for an accuracy that
we would perceive as necessary and sufficient
for, e.~g., reliable microscopic nuclear structure
calculations?

To answer this question, we show
in Table~\ref{tab_chi2a}
the $\chi^2$/datum for the fit of the world $np$ data
below 290 MeV for families of $np$ potentials at 
NLO and NNLO constructed by the Juelich group~\cite{EGM04}.
The NLO potentials produce the very large $\chi^2$/datum between 67 and 105,
and the NNLO are between 12 and 27.
The rate of improvement from one order to the other
is very encouraging, but the quality of the reproduction
of the $np$ data at NLO and NNLO is obviously
insufficient for reliable predictions.

Based upon these facts, it has been pointed out in 2002 by
Entem and Machleidt~\cite{EM02a,EM02} that one has
to proceed to N$^3$LO. Consequently, the first N$^3$LO  potential was
published in 2003~\cite{EM03}.

At N$^3$LO, there are a total of 24 contact terms (24 parameters)
which contribute to the partial waves with $L\leq 2$.
These 24 LECs are essentially free constants
which parametrize
the short-ranged phenomenological part of the interaction.
In Table~\ref{tab_par}, column `$Q^4$/N$^3$LO', we show how these
terms are distributed over the partial waves.
Most important for the improved reproduction of the $NN$ 
phase shifts (and $NN$ observables) 
at N$^3$LO is the fact that 
contacts appear for the first time in $D$-waves. 
$D$-waves are not truely peripheral and, therefore,
1PE plus 2PE alone do not describe them well.
The $D$-wave contacts provide the necessary 
short-range corrections to get the $D$-phases right.
Besides this, at N$^3$LO, another contact is added to each
$P$-wave, which leads to substantial improvements,
particularly, in $^3P_0$ and $^3P_1$ above 100 MeV.

\begin{table}
\caption{Low-energy constants applied in the N$^3$LO $NN$ potential 
(column `$NN$ pot.').
The $c_i$ belong to the dimension-two $\pi N$ Lagrangian
and are in units of GeV$^{-1}$, while the 
$\bar{d}_i$ are associated with the dimension-three Lagrangian
and are in units of GeV$^{-2}$. The column `$\pi N$ analysis' shows values
determined from $\pi N$ data.
\label{tab_LEC}}
\smallskip
\begin{tabular*}{\textwidth}{@{\extracolsep{\fill}}crr}
\hline 
\hline 
\noalign{\smallskip}
 & $NN$ pot. & $\pi N$  analysis\\
\hline
$c_1$ & --0.81 & $-0.81\pm 0.15^a$ \\
$c_2$ & 2.80 & $3.28\pm 0.23^b$ \\
$c_3$ & --3.20 & $-4.69\pm 1.34^a$ \\
$c_4$ & 5.40 & $3.40\pm 0.04^a$ \\
$\bar{d}_1 + \bar{d}_2$ & 3.06 & $3.06\pm 0.21^b$ \\
$\bar{d}_3$ & --3.27 & $-3.27\pm 0.73^b$ \\
$\bar{d}_5$ & 0.45 & $0.45\pm 0.42^b$ \\
$\bar{d}_{14} - \bar{d}_{15}$ & --5.65 & $-5.65\pm 0.41^b$
\\ 
\hline 
\hline 
\end{tabular*}
\footnotesize
$^a$Table~1, Fit~1 of Ref.~\cite{BM00}.\hspace{5mm}
$^b$Table~2, Fit~1 of Ref.~\cite{FMS98}.
\end{table}

In Table~\ref{tab_par},
we also show the number of parameters
used in the Nijmegen partial wave analysis (PWA93)~\cite{Sto93}
and in the high-precision CD-Bonn potential~\cite{Mac01}.
The table reveals that, for $S$ and $P$ waves, 
the number of parameters
used in high-precision phenomenology and in EFT at N$^3$LO
are about the same.
{\bf Thus, the EFT approach provides retroactively a justification
for the phenomenology used in the 1990's to  obtain high-precision fits.}

At NLO and NNLO, the number of contact parameters is substantially
smaller than for PWA93 and CD-Bonn, which explains why
these orders are insufficient for a quantitative potential.
The 24 parameters of N$^3$LO are close to the 30+
used in PWA93 and high precision potentials. Consequently
(see following sections for details), at N$^3$LO,
a fit of the $NN$ data is possible
that is of about the
same quality as the one by the high-precision $NN$ 
potentials~\cite{Sto94,WSS95,MSS96,Mac01}.
Thus, one may perceive N$^3$LO as the order of ChPT that
is necessary and sufficient for a reliable $NN$ potential.

\subsubsection{A quantitative $NN$ potential at N$^3$LO
\label{sec_potn3lo}}

We choose the Idaho N$^3$LO potential~\cite{EM03} as example.
Note that
for an accurate fit of the low-energy $pp$ and $np$ data, 
charge-dependence is important.
We include charge-dependence up to next-to-leading order 
of the isospin-violation scheme 
(NL\O, in the notation of Ref.~\cite{WME01}).
Thus, we include
the pion mass difference in 1PE and the Coulomb potential
in $pp$ scattering, which takes care of the L\O\/ contributions. 
At order NL\O\, we have pion mass difference in the NLO part of TPE,
$\pi\gamma$ exchange~\cite{Kol98}, and two charge-dependent
contact interactions of order $Q^0$ which make possible
an accurate fit of the three different $^1S_0$ scattering 
lengths, $a_{pp}$, $a_{nn}$, and $a_{np}$.

For the cutoff parameter $\Lambda$
of the regulator Eq.~(\ref{eq_regulator}),
we choose initially 500 MeV.
Within a certain reasonable range, results should not
depend sensitively on $\Lambda$ (cf.\ discussion
in Section~\ref{sec_reno}). Therefore, we have also
made a second fit for $\Lambda=600$ MeV.

The fitting procedure starts with 
the peripheral partial waves because they involve
fewer and more fundamental parameters.
Partial waves with $L\geq 3$
are exclusively determined by 1PE and 2PE because
the N$^3$LO contacts contribute to $L\leq 2$ only.
1PE and 2PE at N$^3$LO depend on
the axial-vector coupling constant, $g_A$ (we use $g_A=1.29$),
the pion decay constant, $f_\pi=92.4$ MeV,
and eight low-energy constants (LECs) that appear in the 
dimension-two and dimension-three $\pi N$ Lagrangians.
The LECs are listed in Table~\ref{tab_LEC},
where column `$NN$ pot.' shows the values
used for the present N$^3$LO potential.
In the fitting process,
we varied three of them, namely, $c_2$, $c_3$, and $c_4$.
We found that the other LECs are not very effective in the
$NN$ system and, therefore, we left them at their central
values as determined in $\pi N$ analysis.
The most influential constant is $c_3$, 
which---in terms of magnitude---has to be chosen
on the low side (slightly more than one standard deviation
below its $\pi N$ determination), otherwise there is
too much central attraction.
Concerning $c_4$, our choice $c_4 = 5.4$ GeV$^{-1}$ 
lowers the $^3F_2$ 
phase shift (and slightly the $^1F_3$) bringing it 
into closer agreement with the phase shift 
analysis---as compared to using the $\pi N$ value
$c_4=3.4$ GeV$^{-1}$. The other
$F$ waves and the higher partial waves are essentially unaffected
by this variation of $c_4$. 
Finally, the change of $c_2$ from its $\pi N$ value
of 3.28 GeV$^{-1}$ to 2.80 GeV$^{-1}$ (our choice)
brings about some subtle improvements of the fit, but
it is not essential.
Overall, the fit
of all $J\geq 3$ waves is very good.

\begin{table}[t]
\caption{Columns three to five display the
$\chi^2$/datum for the reproduction of the 1999 
{\boldmath\bf $np$ database}~\cite{note2}
(subdivided into energy intervals)
by various $np$ potentials.
For the chiral potentials, 
the $\chi^2$/datum is stated in terms of ranges
which result from a variation of the cutoff parameters
used in the regulator functions.
The values of these cutoff parameters 
in units of MeV
are given in parentheses.
$T_{\rm lab}$ denotes the kinetic energy of the incident nucleon
in the laboratory system.
\label{tab_chi2b}}
\smallskip
\begin{tabular*}{\textwidth}{@{\extracolsep{\fill}}ccccc}
\hline 
\hline 
\noalign{\smallskip}
 $T_{\rm lab}$ bin
 & \# of {\boldmath $np$}
 & {\it Idaho}
 & {\it Juelich}
 & Argonne         
\\
 (MeV)
 & data
 & N$^3$LO~\cite{EM03}
 & N$^3$LO~\cite{EGM05} 
 & $V_{18}$~\cite{WSS95}
\\
 & 
 & (500--600)
 & (600/700--450/500)
 & 
\\
\hline 
\hline 
\noalign{\smallskip}
0--100&1058&1.0--1.1&1.0--1.1&0.95\\ 
100--190&501&1.1--1.2&1.3--1.8&1.10\\ 
190--290&843&1.2--1.4&2.8--20.0&1.11\\ 
\hline 
\noalign{\smallskip}
0--290&2402&1.1--1.3&1.7--7.9&1.04
\\ 
\hline 
\hline 
\end{tabular*}
\end{table}

\begin{table}
\caption{Same as Table~\ref{tab_chi2b} but for 
{\boldmath\bf $pp$}.
\label{tab_chi2c}}
\smallskip
\begin{tabular*}{\textwidth}{@{\extracolsep{\fill}}ccccc}
\hline 
\hline 
\noalign{\smallskip}
 $T_{\rm lab}$ bin
 & \# of {\boldmath $pp$}
 & {\it Idaho}
 & {\it Juelich}
 & Argonne         
\\
 (MeV)
 & data
 & N$^3$LO~\cite{EM03}
 & N$^3$LO~\cite{EGM05} 
 & $V_{18}$~\cite{WSS95}
\\
 &  
 & (500--600)
 & (600/700--450/500)
 & 
\\
\hline 
\hline 
\noalign{\smallskip}
0--100&795&1.0--1.7&1.0--3.8&1.0 \\ 
100--190&411&1.5--1.9&3.5--11.6&1.3 \\ 
190--290&851&1.9--2.7&4.3--44.4&1.8 \\ 
\hline 
\noalign{\smallskip}
0--290&2057&1.5--2.1&2.9--22.3&1.4 
\\ 
\hline 
\hline 
\end{tabular*}
\end{table}

\begin{figure}[t]
\vspace*{-1.5cm}
\hspace*{-1.7cm}
\scalebox{0.5}{\includegraphics{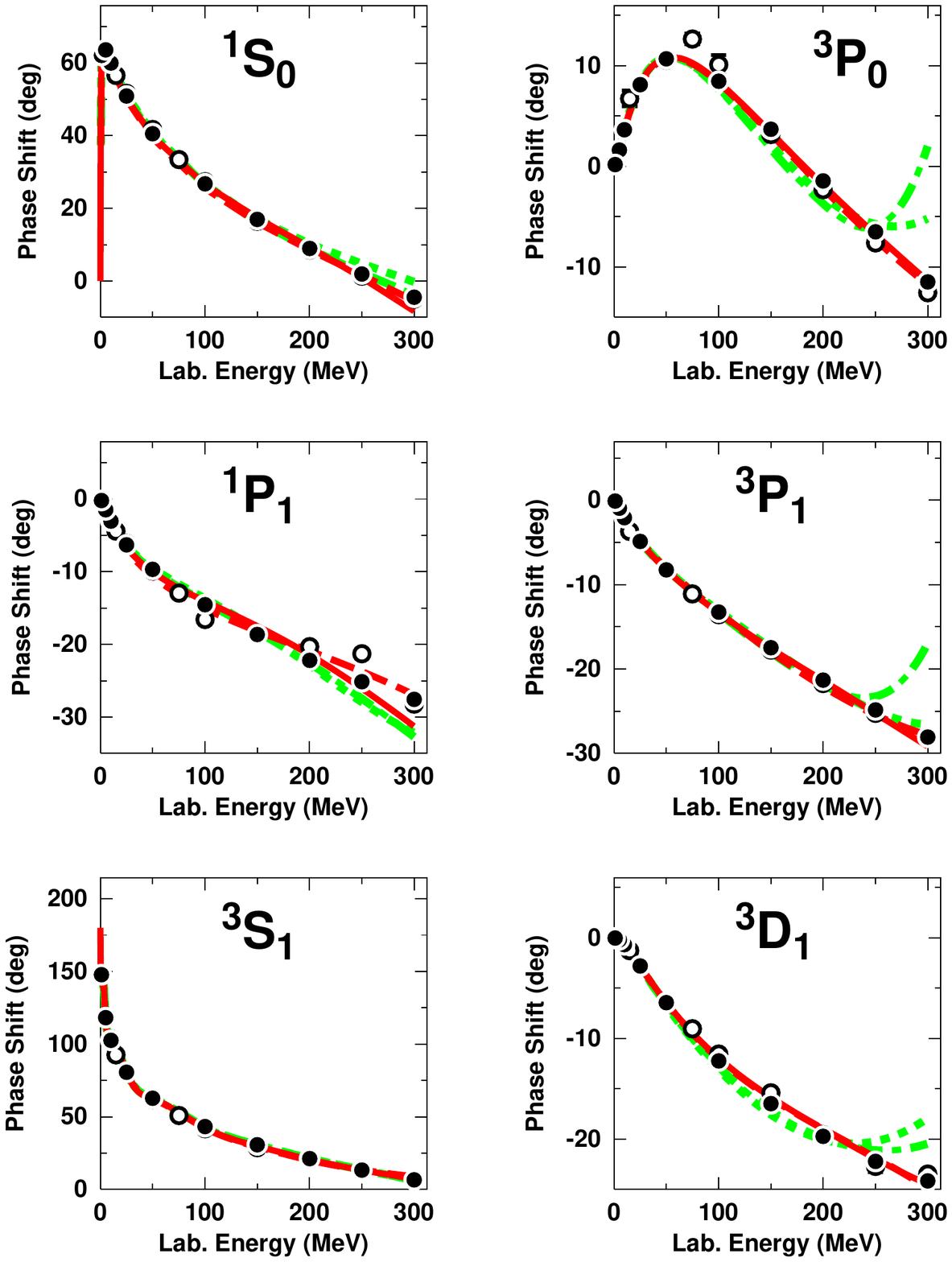}}
\hspace*{-2.5cm}
\scalebox{0.5}{\includegraphics{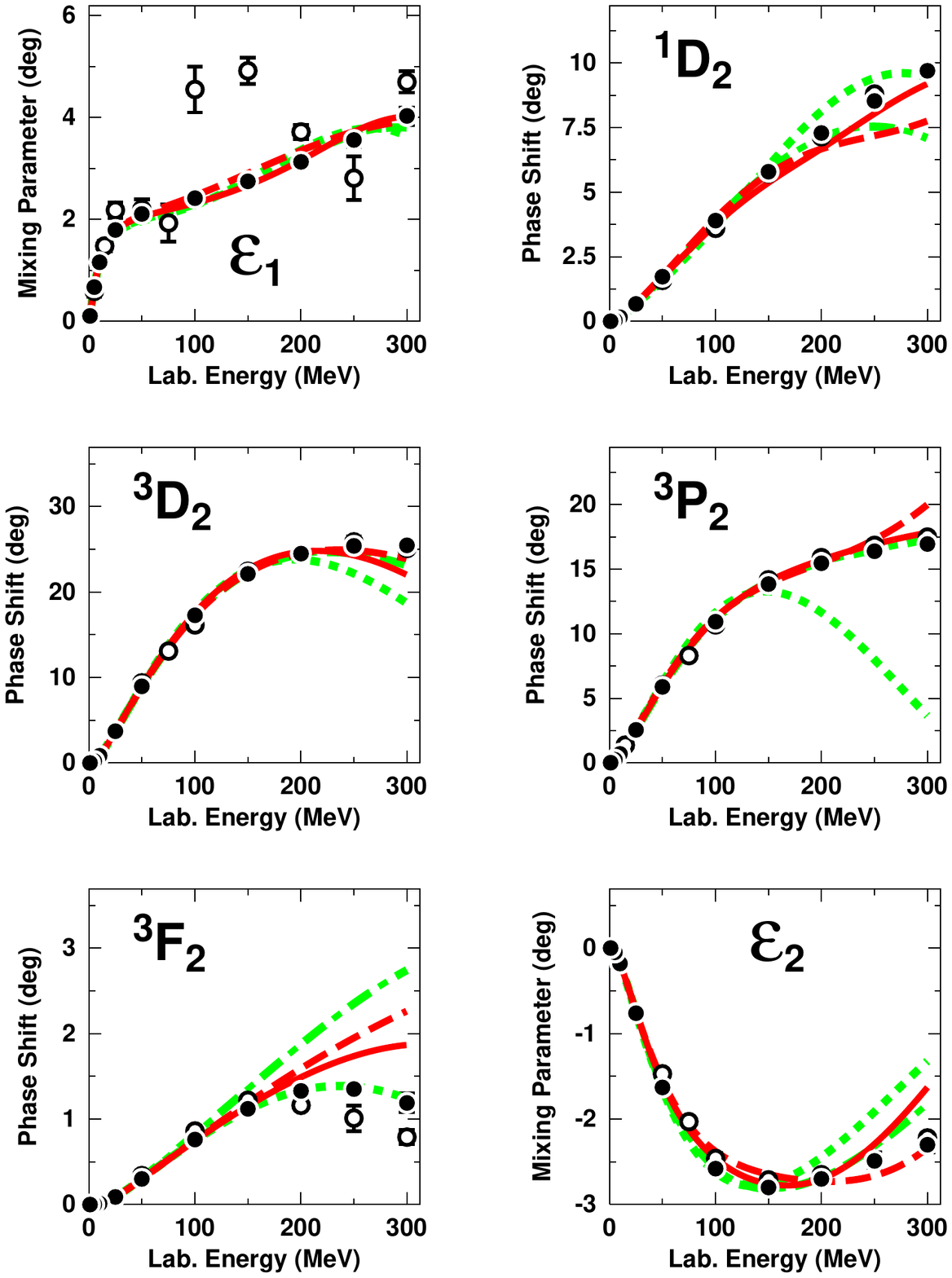}}
\vspace*{-2.5cm}
\caption{Neutron-proton phase parameters as described by
various chiral potentials at N$^3$LO.
The (red) solid and the dashed curves are calculated from
Idaho N$^3$LO potentials~\cite{EM03} with $\Lambda=$ 500 
and 600 MeV, respectively;
while the (green) dash-dotted and the dotted curves are based upon
Juelich N$^3$LO potentials~\cite{EGM05} with cutoff combinations
600/700 and 450/500 MeV, respectively.
Partial waves with total angular momentum $J\leq 2$ and laboratory energies up to 300 MeV are displayed.
The solid dots and open circles are the results from the Nijmegen
multi-energy $np$ phase shift analysis~\protect\cite{Sto93} 
and the VPI/GWU
single-energy $np$ analysis SM99~\protect\cite{SM99}, respectively.
\label{fig_phn3lo1}}
\end{figure}

We turn now to the lower partial waves.
Here, the most important fit parameters are the ones associated
with the 24 contact terms that contribute to the partial waves
with $L\leq 2$.
In addition, we have two charge-dependent
contacts which are used to fit the
three different $^1S_0$ scattering 
lengths, $a_{pp}$, $a_{nn}$, and $a_{np}$.

In the optimization procedure, we fit first phase shifts,
and then we refine the fit by minimizing the
$\chi^2$ obtained from a direct comparison with the data.
We start with $pp$, since the $pp$ phase shifts and data are more
accurate than the $np$ ones. The $pp$ fit fixes essentially the $I=1$ potential.
The $I=1$ $np$ potential is just the $pp$ one modified
by charge-dependence due to nucleon-mass difference, pion-mass splitting in 1PE,
$\pi\gamma$ exchange, and omission of Coulomb. 
In addition to this, the non-derivative contact in the $^1S_0$ state is changed such as to 
reproduce the $np$  scattering length.
The $nn$ potential is the $pp$ one without Coulomb, using neutron masses, and 
fitting the $nn$ scattering length in the $^1S_0$ state with the non-derivative contact.

The $\chi^2/$datum for the fit of the $np$ data below
290 MeV are shown in Table~\ref{tab_chi2b}, 
and the corresponding ones for $pp$
are given in Table~\ref{tab_chi2c}.
These tables reveal that at N$^3$LO
a $\chi^2$/datum comparable to the high-precision
Argonne $V_{18}$~\cite{WSS95} potential can, indeed, be achieved.
The Idaho N$^3$LO potential~\cite{EM03} with $\Lambda=500$ MeV produces
a $\chi^2$/datum = 1.1 
for the world $np$ data below 290 MeV
which compares well with the $\chi^2$/datum = 1.04
by the Argonne potential.
In 2005, also the Juelich group produced
several N$^3$LO $NN$ potentials~\cite{EGM05}, the best of which
fits the $np$ data with
a $\chi^2$/datum = 1.7 and the worse with 
7.9 (Table~\ref{tab_chi2b}).

Turning to $pp$,
the $\chi^2$ for $pp$ data are typically
larger than for $np$
because of the higher precision of $pp$ data.
Thus, the Argonne $V_{18}$ produces
a $\chi^2$/datum = 1.4 for the world $pp$ data
below 290 MeV and the best Idaho N$^3$LO $pp$ potential obtains
1.5. The fit by the best Juelich 
N$^3$LO $pp$ potential results in
a $\chi^2$/datum = 2.9 and the worst  produces 22.3.

Phase shifts of $np$ scattering from two Idaho 
(solid and dashed lines) and two Juelich (dash-dotted 
and dotted lines)
N$^3$LO $np$ potentials are shown in 
Figs.~\ref{fig_phn3lo1}.
The phase shifts confirm what the corresponding
$\chi^2$ have already revealed.
For the low-energy scattering and deuteron parameters, see Ref.~\cite{EM03}.

\section{Few-nucleon forces \label{sec_manyNF}}

The chiral 2NF discussed in the previous section has been applied in microscopic
calculations of nuclear structure with, in general, a great deal of 
success~\cite{Cor02,Cor05,Cor10,NC04,FNO05,Var05,Kow04,DH04,Wlo05,Dea05,Gou06,Hag08,Hag10,FOS04,FOS09}.
However, from high-precision studies conducted in the 1990s, it is well-known that certain few-nucleon
reactions and nuclear structure issues require 3NFs for their microscopic explanation.
Outstanding examples are the $A_y$ puzzle of $N$-$d$ scattering~\cite{Glo96,EMW02}
and the ground state of $^{10}$B~\cite{Cau02}.
As noted before, 
an important advantage of the EFT approach to nuclear forces
is that it creates two- and many-nucleon forces on an equal
footing (cf.\ the overview given in Fig.~\ref{fig_hi}).

For a 3NF, we have $A=3$ and $C=1$ and, thus, Eq.~(\ref{eq_nu})
implies 
\begin{equation}
\nu = 2 + 2L + 
\sum_i \Delta_i \,.
\label{eq_nu3nf}
\end{equation}
This equation can be used to analyze 3NF contributions
order by order.
The lowest possible power is obviously $\nu=2$ (NLO), which
is obtained for no loops ($L=0$) and 
only leading vertices
($\sum_i \Delta_i = 0$). 
This 3NF happens to vanish~\cite{Wei92,YG86,CF86}.
The first non-vanishing 3NF occurs at NNLO.

\begin{figure}[t]\centering
\vspace*{-4.0cm}
\scalebox{0.55}{\includegraphics{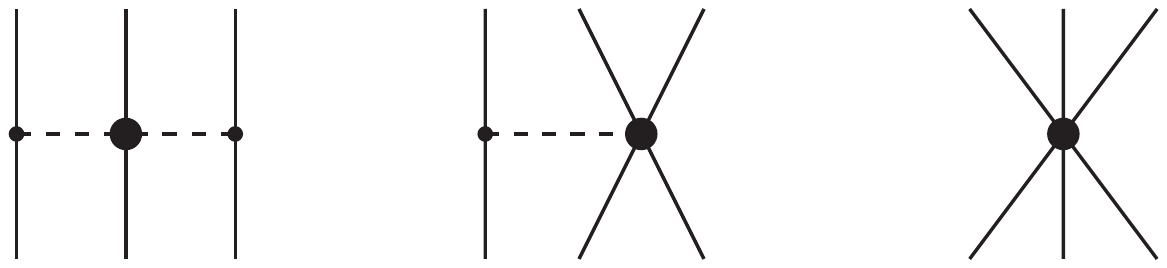}}
\vspace*{-10.0cm}
\caption{The three-nucleon force at NNLO.
From left to right: 2PE, 1PE, and contact diagrams.
Notation as in Fig.~\ref{fig_hi}.}
\label{fig_3nf_nnlo}
\end{figure}

The power $\nu=3$ (NNLO) is obtained when
there are no loops ($L=0$) and 
$\sum_i \Delta_i = 1$, i.e., 
$\Delta_i=1$ for one vertex 
while $\Delta_i=0$ for all other vertices.
There are three topologies which fulfill this condition,
known as the two-pion exchange (2PE), 1PE,
and contact graphs~\cite{Kol94,Epe02b}
(Fig.~\ref{fig_3nf_nnlo}).

The 2PE 3N-potential is given by
\begin{equation}
V^{\rm 3NF}_{\rm 2PE} = 
\left( \frac{g_A}{2f_\pi} \right)^2
\frac12 
\sum_{i \neq j \neq k}
\frac{
( \vec \sigma_i \cdot \vec q_i ) 
( \vec \sigma_j \cdot \vec q_j ) }{
( q^2_i + m^2_\pi )
( q^2_j + m^2_\pi ) } \;
F^{ab}_{ijk} \;
\tau^a_i \tau^b_j
\label{eq_3nf_nnloa}
\end{equation}
with $\vec q_i \equiv \vec{p_i}' - \vec p_i$, 
where 
$\vec p_i$ and $\vec{p_i}'$ are the initial
and final momenta of nucleon $i$, respectively, and
\begin{equation}
F^{ab}_{ijk} = \delta^{ab}
\left[ - \frac{4c_1 m^2_\pi}{f^2_\pi}
+ \frac{2c_3}{f^2_\pi} \; \vec q_i \cdot \vec q_j \right]
+ 
\frac{c_4}{f^2_\pi}  
\sum_{c} 
\epsilon^{abc} \;
\tau^c_k \; \vec \sigma_k \cdot [ \vec q_i \times \vec q_j] \; .
\label{eq_3nf_nnlob}
\end{equation}  
There are great similarities between this force and earlier derivations of
2PE 3NFs, notably the 50-year old Fujita-Miyazawa~\cite{FM57},
the Tucson-Melbourne (TM)~\cite{Coo79}, and the Brazil~\cite{CDR83}
forces. This is also the 3NF advocated by Gerry Brown in Refs.~\cite{BGG68,Bro79}.

The 1PE contribution is
\begin{equation}
V^{\rm 3NF}_{\rm 1PE} = 
-D \; \frac{g_A}{8f^2_\pi} 
\sum_{i \neq j \neq k}
\frac{\vec \sigma_j \cdot \vec q_j}{
 q^2_j + m^2_\pi }
( \mbox{\boldmath $\tau$}_i \cdot \mbox{\boldmath $\tau$}_j ) 
( \vec \sigma_i \cdot \vec q_j ) 
\label{eq_3nf_nnloc}
\end{equation}
and the 3N contact potential reads
\begin{equation}
V^{\rm 3NF}_{\rm ct} = E \; \frac12
\sum_{j \neq k} 
 \mbox{\boldmath $\tau$}_j \cdot \mbox{\boldmath $\tau$}_k  \; .
\label{eq_3nf_nnlod}
\end{equation}
The last two 3NF terms involve the two new parameters $D$ and $E$, 
which do not appear in the 2N problem.
There are many ways to pin these two parameters down.
In Ref.~\cite{Epe02b},
the triton binding energy and the $nd$ doublet scattering
length $^2a_{nd}$ were used.
One may also choose the binding
energies of $^3$H and $^4$He~\cite{Nog06} or
an optimal over-all fit of the properties of light nuclei~\cite{Nav07}.
Once $D$ and $E$ are fixed, the results for other
3N, 4N, etc.\  observables are predictions.

The 3NF at NNLO has been applied in
calculations of few-nucleon 
reactions~\cite{Epe02b,Erm05,Kis05,Wit06,Ley06,Ste07,KE07,Mar09,Kie10,Viv10},
structure of light- and medium-mass 
nuclei~\cite{Nog06,Nav07,Hag07,Ots09},
and nuclear and neutron matter~\cite{Bog05,HS09}
with a good deal of success.
Yet, the famous `$A_y$ puzzle' of nucleon-deuteron scattering
is not resolved~\cite{Epe02b,KE07}.
When only 2NFs are applied,
the analyzing power in $p$-$^3$He scattering
is even more underpredicted than in $p$-$d$~\cite{DF07,Fis06}. 
However, when the
NNLO 3NF is added, the $p$-$^3$He $A_y$ substantially 
improves (more than in $p$-$d$)~\cite{Viv10}---but a discrepancy remains.
Furthermore, the spectra of light nuclei leave room for improvement~\cite{Nav07}.

To summarize, the 3NF at NNLO is a remarkable contribution: It represents the leading many-body force
within the scheme of ChPT; it includes terms that were advocated by Gerry Brown already some 40 years 
ago~\cite{BGG68,Bro79}; and it
produces noticeable improvements in few-nucleon reactions and the structure of
light nuclei.
But unresolved problems remain. Moreover, in the case of the
2NF, we have seen that one has to proceed to N$^3$LO to achieve sufficient accuracy.
Therefore, the 3NF at N$^3$LO is needed for at least two reasons: 
for consistency with the 2NF and to hopefully resolve outstanding problems in microscopic structure
and reactions. For further discussion of the 3NF (beyond NNLO), see Ref.~\cite{ME10}.

\section{Conclusions and Outlook
\label{sec_concl}}

Forty years after Gerry Brown had stressed repeatedly the outstanding importance of
chiral symmetry, we have finally arrived at a derivation of nuclear forces which takes this symmetry
consistently into account.

The greatest progress occurred during the past 15 years.
Key to this development was the realization that
low-energy QCD is equivalent to a chiral effective field theory which allows for 
a perturbative expansion that has become known as chiral perturbation theory (ChPT).
In this framework, two- and many-body forces emerge on an equal footing and the empirical fact
that nuclear many-body forces are substantially weaker than the two-nucleon force
is explained naturally.

In this chapter, we have shown how the two-nucleon force is derived
from ChPT and demonstrated that,
at N$^3$LO, the accuracy can be achieved that
is necessary and sufficient for reliable microscopic nuclear 
structure predictions.
First calculations applying the N$^3$LO
$NN$ potential~\cite{EM03}
in the conventional shell model~\cite{Cor02,Cor05,Cor10},
the {\it ab initio} no-core shell model~\cite{NC04,FNO05,Var05},
the coupled cluster formalism~\cite{Kow04,DH04,Wlo05,Dea05,Gou06,Hag08,Hag10},
and the unitary-model-operator approach~\cite{FOS04,FOS09}
have produced promising results. 

We also discussed nuclear many-body forces based upon chiral EFT.
The 3NF at NNLO has been known for a while~\cite{Kol94,Epe02b} and
applied in 
few-nucleon reactions~\cite{Epe02b,Erm05,Kis05,Wit06,Ley06,Ste07,KE07,Mar09,Kie10,Viv10},
structure of light- and medium-mass nuclei~\cite{Nog06,Nav07,Hag07,Ots09},
and nuclear and neutron matter~\cite{Bog05,HS09} with good success;
but some open issues remain, which are presently under investigation~\cite{ME10}.

We are optimistic that
the remaining outstanding problems will be resolved within the next few years, such that,
after 80 years of desperate struggle, we
may finally claim that the nuclear force problem is essentially under 
control---last not least, due to a thorough accounting of chiral symmetry.

\section*{Acknowledgement}
It is our great pleasure to dedicate this chapter to Gerry Brown on the occasion of his 85th birthday. 
One of us (R.M.) has been a postdoc with Gerry during the years of 1976 and 1977 and stayed in contact
with him ever since. R.M. has always been impressed by Gerry's liberating personality and 
inspiring physics intuition.
This work was supported in part by the U.S. Department of Energy
under Grant No.~DE-FG02-03ER41270 (R.M.), and the Ministerio de Ciencia y
Tecnolog\'\i a under Contract No.~FPA2007-65748 and the Junta de Castilla
y Le\'on under Contract Nos.~SA-106A07 and GR12 (D.R.E.).


\begin{thebibliography}{99}
\bibitem{Bro70}
G. E. Brown, Comments Nucl. Part. Phys. {\bf 4} (1970) 140.
\bibitem{Bro79}
G. E. Brown, in: Meson in Nuclei, Vol. I, edited by M. Rho and D. H. Wilkinson
(North-Holland, Amsterdam, 1979) p.~330.
\bibitem{BGG68}
G. E. Brown, A. M. Green, and W J. Gerace, Nucl. Phys. {\bf A115} (1968) 435.

\bibitem{Sch57} J. Schwinger, Ann. Phys. (N.Y.) {\bf 2} (1957) 407.
\bibitem{GL60} M. Gell-Mann and M. Levy, Nuovo Cimento {\bf 16} (1960) 705.
\bibitem{EW88} T. Ericson and W. Weise, {\it Pions and Nuclei} (Clarendon Press, Oxford, 1988).
\bibitem{Wei67} S. Weinberg, Phys. Rev. Lett. {\bf 18} (1967) 188.
\bibitem{Wei66} S. Weinberg, Phys. Rev. Lett. {\bf 17} (1966) 616.
\bibitem{Tom66}
Y. Tomozawa, Nuovo Cimento A {\bf 46} (1966) 707.
\bibitem{Wei68} S. Weinberg, Phys. Rev. {\bf 166} (1968) 1568.
\bibitem{CCWZ} 
S. Coleman, J. Wess, and B. Zumino, Phys.\ Rev.\ {\bf 177} (1969) 2239;
C. G. Callan, S. Coleman, J. Wess, and B. Zumino, 
{\it ibid.} {\bf 177} (1969) 2247.
\bibitem{Wei79} S. Weinberg, Physica {\bf 96A} (1979) 327.
\bibitem{Wei09}
S. Weinberg, {\it Effective Field Theory, Past and Future},
arXiv:0908:1964 [hep-th].
\bibitem{GL84} J. Gasser and H. Leutwyler,
Ann.\ Phys.\ {\bf 158} (1984) 142; 
Nucl. Phys. {\bf B250} (1985) 465.
\bibitem{GSS88} J. Gasser, M. E. Sainio, and A. \v{S}varc,
Nucl.\ Phys.\ {\bf B307} (1986) 779.
\bibitem{Wei90} S. Weinberg, Phys.\ Lett.\ B {\bf 251} (1990) 288.
\bibitem{Wei91} S. Weinberg, Nucl.\ Phys.\ {\bf B363} (1991) 3.
\bibitem{Wei92} S. Weinberg, Phys.\ Lett.\ B {\bf 295} (1992) 114.
\bibitem{ORK94}
C. Ord\'o\~nez, L. Ray, and U. van Kolck,
Phys.\ Rev.\ Lett.\ {\bf 72} (1994) 1982.
\bibitem{ORK96}
C. Ord\'o\~nez, L. Ray, and U. van Kolck,
Phys.\ Rev.\ C {\bf 53} (1996) 2086.
\bibitem{KBW97} N. Kaiser, R. Brockmann, and W. Weise,
Nucl.\ Phys.\ {\bf A625} (1997) 758.
\bibitem{KGW98} N. Kaiser, S. Gerstend\"orfer, and W. Weise,
Nucl.\ Phys.\ {\bf A637} (1998) 395.
\bibitem{Kai00a} N. Kaiser, Phys.\ Rev.\ C {\bf 61} (2000) 014003.
\bibitem{Kai00b} N. Kaiser, Phys.\ Rev.\ C {\bf 62} (2000) 024001.
\bibitem{Kai01} N. Kaiser, Phys.\ Rev.\ C {\bf 63} (2001) 044010.
\bibitem{Kai01a} N. Kaiser, Phys.\ Rev.\ C {\bf 64} (2001) 057001.
\bibitem{Kai01b} N. Kaiser, Phys. Rev. C {\bf 65} (2002) 017001.
\bibitem{RR94}
C. A. da Rocha and M. R. Robilotta, 
Phys.\ Rev.\ C {\bf 49} (1994) 1818;
{\bf 52} (1995) 531;
M. R. Robilotta, Nucl. Phys. {\bf A595} (1995) 171;
M. R. Robilotta and C. A. da Rocha, 
Nucl. Phys. {\bf A615} (1997) 391;
J.-L. Ballot, C. A. da Rocha, and M. R. Robilotta, 
Phys. Rev. C {\bf 57} (1998) 1574.
\bibitem{RR03}
R. Higa and M. R. Robilotta, 
Phys. Rev. C {\bf 68} (2003) 024004.
\bibitem{EGM98}
E. Epelbaum, W. Gl\"ockle, and U.-G. Mei\ss ner, 
Nucl.\ Phys.\ {\bf A637} (1998) 107.
\bibitem{EGM00} 
E. Epelbaum, W. Gl\"ockle, and U.-G. Mei\ss ner,
Nucl. Phys. {\bf A671} (2000) 295.
\bibitem{EM02a} D. R. Entem and R. Machleidt,
Phys.\ Lett.\ B {\bf 524} (2002) 93.
\bibitem{EM02} D. R. Entem and R. Machleidt,
Phys. Rev. C {\bf 66} (2002) 014002.
\bibitem{EM03} D. R. Entem and R. Machleidt,
Phys.\ Rev.\ C {\bf 68} (2003) 041001.
\bibitem{ME05}  R. Machleidt and D. R. Entem,
J.\ Phys.\ G: Nucl.\ Phys.\ {\bf 31} (2005) S1235.
\bibitem{Kol94} U. van Kolck, Phys. Rev. C {\bf 49} (1994) 2932.
\bibitem{Epe02b} E. Epelbaum {\it et al.}, 
Phys. Rev. C {\bf 66} (2002) 064001.
\bibitem{IR07} 
S. Ishikawa and M. R. Robilotta, 
{ Phys. Rev.} C {\bf 76} (2007) 014006.
\bibitem{Ber08}
V. Bernard, E. Epelbaum, H. Krebs, and U.-G. Mei\ss ner,
{ Phys. Rev.} C {\bf 77} (2008) 064004.

\bibitem{DF07}
A. Deltuva and A. C. Fonseca, 
Phys. Rev. Lett. {\bf 98} (2007) 162502;
Phys. Rev. C {\bf 76}, (2007) 021001;
arXiv:1005.1308 [nucl-th].

\bibitem{Cor02} L. Coraggio {\it et al.}, Phys.\ Rev.\ C {\bf 66} (2002)
021303.
\bibitem{Cor05} L. Coraggio {\it et al.}, Phys.\ Rev.\ C {\bf 71} (2005)
014307.
\bibitem{Cor10} L. Coraggio {\it et al.}, arXiv:1005.2896 [nucl-th].
\bibitem{NC04} P. Navr\'atil and E. Caurier, {\it Phys.\ Rev.\ C}
{\bf 69} (2004) 014311.
\bibitem{FNO05} C. Forssen {\it et al.},
Phys.\ Rev.\ C {\bf 71} (2005) 044312.
\bibitem{Var05} J.P. Vary {\it et al.}, Eur.\ Phys.\ J.\ A
{\bf 25} s01 (2005) 475.
\bibitem{Kow04} K. Kowalski {\it et al.}, 
Phys.\ Rev.\ Lett.\ {\bf 92} (2004) 132501.
\bibitem{DH04} D.J. Dean and M. Hjorth-Jensen,
{\it Phys. Rev.} C {\bf 69} (2004) 054320.
\bibitem{Wlo05} M. Wloch {\it et al.}, 
J.\ Phys.\ G {\bf 31} (2005) S1291;
Phys.\ Rev.\ Lett.\ {\bf 94} (2005) 21250.
\bibitem{Dea05} D.J. Dean {\it et al.}, Nucl.\ Phys.\
{\bf 752} (2005) 299.
\bibitem{Gou06} J.R. Gour {\it et al.}, Phys.\ Rev.\ C
{\bf 74} (2006) 024310.
\bibitem{Hag08}
G. Hagen, T. Papenbrock, D. J. Dean, and M. Hjorth-Jensen,
Phys. Rev. Lett. {\bf 101} (2008) 092502.
\bibitem{Hag10}
G. Hagen, T. Papenbrock, D. J. Dean, and M. Hjorth-Jensen,
Phys. Rev. C {\bf 82}, 034330 (2010).
\bibitem{FOS04} S. Fujii, R. Okamoto, and K. Suzuki,
Phys.\ Rev.\ C {\bf 69} (2004) 034328.
\bibitem{FOS09} S. Fujii, R. Okamoto, and K. Suzuki,
Phys. Rev. Lett. {\bf 103} (2009) 182501.
\bibitem{Erm05} K. Ermisch {\it et al.}, Phys.\ Rev.\ C
{\bf 71} (2005) 064004.
\bibitem{Kis05}
S. Kistryn {\it et al.}, Phys. Rev. C {\bf 72} (2005) 044006.
\bibitem{Wit06} H. Witala, J. Golak, R. Skibinski, W. Gl\"ockle, A. Nogga,
E. Epelbaum, H. Kamada, A. Kievsky, and M. Viviani,
Phys.\ Rev.\ C {\bf 73} (2006) 044004.
\bibitem{Ley06}
J. Ley {\it et al.}, Phys. Rev. C {\bf 73} (2006) 064001.
\bibitem{Ste07}
E. Stephan {\it et al.}, Phys. Rev. C {\bf 76} (2007) 057001.
\bibitem{KE07}
N. Kalantar-Nayestanaki and E. Epelbaum, 
{\it The three-nucleon system as a laboratory for nuclear physics:
the need for 3N forces}, Nucl. Phys. News {\bf 17} (2007) 22,
arXiv:nucl-th/0703089, and references therein.
\bibitem{Mar09}
L. C. Marcucci, A. Kievsky, L. Girlanda, S. Rosati, and M. Viviani,
{ Phys. Rev.} C {\bf 80} (2009) 034003.
\bibitem{Kie10}
A. Kievsky, M. Viviani, L. Girlanda, and L. E. Marcucci,
Phys. Rev. C {\bf 81} (2010) 044003.
\bibitem{Viv10}
M. Viviani, L. Giarlanda, A. Kievsky, L. E. Marcucci, and S. Rosati,
arXiv:1004.1306 [nucl-th].

\bibitem{Nog06}
A. Nogga, P. Navratil, B. R. Barrett, and J. P. Vary,
Phys.\ Rev.\ C {\bf 73} (2006) 064002.
\bibitem{Nav07}
P. Navratil, V. G. Gueorguiev, J. P. Vary, W. E. Ormand,
and A. Nogga, Phys. Rev. Lett. {\bf 99} (2007) 042501.
\bibitem{Hag07}
G. Hagen {\it et al}, { Phys. Rev.} C {\bf 76} (2007) 034302.
\bibitem{Ots09}
T. Otsuka, T. Susuki, J. D. Holt, A. Schwenk, and Y. Akaishi,
Phys. Rev. Lett. {\bf 105} (2010) 03250.
\bibitem{Bog05} 
S. K. Bogner {\it et al.}, Nucl. Phys. {\bf A763} (2005) 59;
S. K. Bogner {\it et al.}, 
{\it Nuclear matter from chiral low-momentum interactions},
arXiv:0903.3366 [nucl-th].
\bibitem{HS09}
K. Hebeler and A. Schwenk,
Phys. Rev. C {\bf 82}, 014314 (2010).

\bibitem{Bea06}
S. R. Beane, P. F. Bedaque, K. Orginos, and M. J. Savage,
Phys. Rev. Lett. {\bf 97} (2006) 012001.
\bibitem{IAH07}
N Ishii, S. Aoki, and T. Hatsuda,
Phys. Rev. Lett. {\bf 99} (2007) 022001.

\bibitem{Gol61}
J. Goldstone, Nuovo Cim. {\bf 19} (1961) 154.
\bibitem{GSW62}
J. Goldstone, A. Salam, and S. Weinberg, Phys. Rev. {\bf 127} (1962) 965.

\bibitem{MHE87} R. Machleidt, K. Holinde, and Ch.\ Elster,
Phys.\ Rep.\ {\bf 149} (1987) 1.
\bibitem{Lac80} 
M. Lacombe, B. Loiseau, J. M. Richard, R. Vinh Mau,
J. C\^{o}t\'{e}, P. Pires, and R. de Tourreil, 
Phys. Rev. C {\bf 21} (1980) 861.
\bibitem{YG86} S. N. Yang and W. Gl\"ockle,
Phys. Rev. C {\bf 33} (1986) 1774.
\bibitem{CF86} S. A. Coon and J. L. Friar,
Phys. Rev. C {\bf 34} (1986) 1060.

\bibitem{Mac89} R. Machleidt, Adv. Nucl. Phys. {\bf 19} (1989) 189.

\bibitem{KSW96}
D. B. Kaplan, M. J. Savage, and M. B. Wise,  
Nucl. Phys. {\bf B478} (1996) 629;
Phys. Lett. B {\bf 424} (1998) 390;
Nucl. Phys. {\bf B534} (1998) 329.
\bibitem{ME10}
R. Machleidt and D. R. Entem, J. Phys. G: Nucl. Part. Phys. {\bf 37}, 064041 (2010).
\bibitem{NTK05}
A. Nogga, R. G. E. Timmermans, and U. van Kolck,
Phys. Rev. C {\bf 72} (2005) 054006.
\bibitem{Val09}
M. P. Valderrama,
Phys. Rev. C {\bf 83}, 024003 (2011).
\bibitem{Mac09}
R. Machleidt, P. Liu, D. R. Entem, and E. Ruiz Arriola,
Phys. Rev. C {\bf 81} (2010) 024001.
\bibitem{Lep97}
G. P. Lepage, 
{\it How to Renormalize the Schr\"odinger Equation},
nucl-th/9706029.
\bibitem{EGM05} E. Epelbaum, W. Gl\"ockle, and U.-G. Mei\ss ner,
Nucl.\ Phys.\ {A747} (2005) 362.


\bibitem{note2} The 1999 $NN$ data base is defined in 
Ref.~\cite{Mac01}.
\bibitem{Mac01} R. Machleidt, Phys. Rev. C {\bf 63} (2001) 024001.

\bibitem{EGM04} E. Epelbaum, W. Gl\"ockle, and U.-G. Mei\ss ner,
Eur.\ Phys.\ J.\ {\bf A19} (2004) 401.

\bibitem{Sto93} V.\ G.\ J.\ Stoks, R.\ A.\ M.\ Klomp, 
M.\ C.\ M.\ Rentmeester, and J.\ J.\ de Swart, 
Phys.\ Rev.\ C {\bf 48} (1993) 792.

\bibitem{Sto94} 
V.\ G.\ J.\ Stoks, R.\ A.\ M.\ Klomp, 
C.\ P.\ F.\ Terheggen, and J.\ J.\ de Swart, 
Phys.\ Rev.\ C {\bf 49} (1994) 2950.
\bibitem{WSS95} R.\ B.\ Wiringa, V.\ G.\ J.\ Stoks, and R. Schiavilla,
Phys.\ Rev.\ C {\bf 51} (1995) 38.
\bibitem{MSS96}
R. Machleidt, F. Sammarruca, and Y. Song,
Phys. Rev. C {\bf 53} (1996) 1483.

\bibitem{WME01} M. Walzl {\it et al.},
Nucl. Phys. {\bf A693}, 663 (2001).
\bibitem{Kol98} U. van Kolck {\it et al.},
Phys. Rev. Lett. {\bf 80} (1998) 4386.

\bibitem{BM00} P. B\"{u}ttiker {\it et al.},
Nucl.\ Phys.\ {\bf A668}, 97 (2000).
\bibitem{FMS98} N. Fettes {\it et al.},
Nucl.\ Phys.\ {\bf A640}, 199 (1998).

\bibitem{SM99} 
R. A. Arndt, I. I. Strakovsky, and R. L. Workman,
SAID, Scattering Analysis Interactive Dial-in computer facility,
George Washington University
(formerly Virginia Polytechnic Institute), 
solution SM99 (Summer 1999); for more information see, e.~g.,
R. A. Arndt, I. I. Strakovsky, and R. L. Workman,
Phys. Rev. C {\bf 50} (1994) 2731.

\bibitem{Glo96}
W. Gl\"ockle {\it et al.}, Phys. Rep. {\bf 274} (1996) 107.
\bibitem{EMW02} D. R. Entem, R. Machleidt, and H. Witala,
Phys. Rev. C {\bf 65} (2002) 064005, and references to earlier work therein.
\bibitem{Cau02}
E. Caurier {\it et al.}, Phys. Rev. C {\bf 66} (2002) 024314.

\bibitem{FM57}
J.-I. Fujita and H. Miyazawa, Prog. Theor. Phys. {\bf 17} (1957) 360. 
\bibitem{Coo79}
S. A. Coon, M. D. Scadron, P. C. McNamee, B. R. Barrett, D. W. E. Blatt. and B. H. J. McKeller,
Nucl. Phys. {\bf A317} (1979) 242.
\bibitem{CDR83}
H. T. Coelho, T. K. Das, and M. R. Robilotta,
Phys. Rev. C {\bf 28} (1983) 1812;
M. R. Robilotta and H. T. Coelho,
Nucl. Phys. {\bf A460} (1986) 645.

\bibitem{Fis06}
B. M. Fisher {\it et al.}, 
Phys. Rev. C {\bf 74} (2006) 034001.

\end{thebibliography}
\end{document}